\documentclass[aps,amsmath,twocolumn]{revtex4}
\usepackage[dvips]{graphicx}           
\usepackage{epsfig}
\usepackage{graphics}
\usepackage{amsmath}
\usepackage{xspace}
\usepackage{pstricks,pst-plot,pst-gr3d,pst-node,epsfig}

\newcommand{\ie}{{\it i.e.}\xspace}

\newcommand{\elabel}[1]{\label{eq:#1}}
\newcommand{\eref}[1]{Eq.~(\ref{eq:#1})}

\newcommand{\eg}{{\it e.g.}\xspace}

\newcommand{\ave}[1]{\left\langle #1 \right \rangle}
\newcommand{\abs}[1]{|#1|}
\newcommand{\flabel}[1]{\label{fig:#1}}
\newcommand{\fref}[1]{Fig.~\ref{fig:#1}}

\begin{document}
\title{Universality under conditions of self-tuning} 

\author{Ole Peters$^{1,2}$} 
\email{ole@santafe.edu}
\author{Michelle Girvan$^{2,3}$}
\email{girvan@umd.edu}
\affiliation{$^1$Department of Mathematics and Grantham Institute for Climate Change,
Imperial College London, 180 Queen's Gates,
London SW7 2 AZ, UK
\\
$^2$Santa Fe Institute,
1399 Hyde Park Road, Santa Fe, NM, 87501, USA \\ 
$^3$University of Maryland, IREAP, Bldg 223, Paint Branch Dr., College Park, MN 20742
}

\begin{abstract}
We study systems with a continuous phase transition that tune their
parameters to maximize a quantity that diverges solely at a unique
critical point. Varying the size of these systems with dynamically
adjusting parameters, the same finite-size scaling is observed as in
systems where all relevant parameters are fixed at their critical
values. This scheme is studied using a self-tuning variant of the
Ising model. It is contrasted with a scheme where systems approach
criticality through a target value for the order parameter that
vanishes with increasing system size.  In the former scheme, the
universal exponents are observed in na\"ive finite-size scaling
studies, whereas in the latter they are not.
\end{abstract}

\begin{sloppypar}
\maketitle

One historical motivation for the study of critical phenomena are the
observable effects of diverging correlation lengths, such as critical
opalescence \cite{Andrews1869}. On a more fundamental level,
universality, the fact that a variety of phenomenologically different
systems share the exact same critical behavior, reveals a deeply
engrained mathematical structure in physical systems.

Universality is well understood in equilibrium systems where
renormalization group methods can be applied. Far-from-equilibrium
systems, often only described by dynamical rules, do not always lend
themselves to the same methods of analysis, and as a consequence the
understanding of universality is less complete. Some systems add
another complication: they self-tune.  We ask what happens to
universality under conditions of self-tuning. In particular, we
investigate in this paper a self-tuning mechanism that reproduces the
universal finite size scaling of thermodynamic observables.  While we
investigate these issues in a near-equilibrium system, the arguments
we put forward may well be applicable to the far-from-equilibrium
systems typically studied in the literature on self-organized
criticality (SOC).

The term self-organized criticality has been used with many different
meanings in different disciplines, often simply to describe a system
whose internal dynamics lead to a degree of scale freedom in global
observables. A more specific definition of SOC, which we employ here,
is the spontaneous emergence of critical behavior in systems with
continuous phase transitions. For instance, sandpile models have been
described in these
terms~\cite{TangBak1988,FraysseSornetteSornette1993,DickmanVespignaniZapperi1998}.
In these systems, defined on $d$-dimensional lattices, local rules
demand the toppling of particles to neighboring sites whenever a
threshold value of the local particle density is exceeded. The
boundaries are open such that particles can be dissipated, and a slow
drive is implemented as an addition of a particle whenever the system
reaches a globally stable state (no supercritical local particle
densities). In these systems the distribution of avalanche sizes,
defined as the number of local reconfigurations in response to the
addition of a particle, is scale-free. Moments of the distribution
show simple finite-size scaling \cite{Manna1991}, just like moments of
the order-parameter distribution in equilibrium critical phenomena
\cite{PrivmanHohenbergAharony1991}.

In 1988 Tang and Bak linked their sandpile model to ordinary
non-equilibrium phase transitions~\cite{TangBak1988}. The overall
particle density, $\zeta$, was identified as the tuning parameter and
the density of active sites, $\rho_a$, also called the ``activity'',
as the order parameter. Both are common observables in continuous
phase transitions, and their identification enables the use of the
powerful formalism of critical phenomena. Investigations of avalanche
size distributions, which are characteristic of the smaller body of
literature on SOC, have been developed less extensively. We use a
notation inspired by absorbing-state (AS) phase transitions
\cite{DickmanVespignaniZapperi1998}, for a review see
\cite{DickmanETAL2000} and references therein. Below a critical value,
$\zeta_c$, of the tuning parameter, the order parameter tends to zero
since local thresholds are rarely surpassed anywhere in the system and
hardly any topplings occur. The order parameter shows very good
finite-size scaling, identical to that of corresponding AS phase
transitions. In these corresponding models the boundaries are closed,
and one measures quasi-stationary values of activities at fixed
particle densities,
$\zeta$~\cite{DickmanETAL2001,ChristensenETAL2004}.  In other words,
there is ample numerical evidence supporting the fact that standard
observables such as the order parameter in sandpile models respect the
universality classes of their corresponding phase transitions. This is
also reflected in the observation that avalanche-size exponents are
directly related to the scaling exponents describing the order
parameter, correlation length, and survival time distribution in the
corresponding AS systems \cite{Luebeck2004}.

Sandpiles are defined in terms of their microscopic dynamics. It is
desirable to mirror the effect of these dynamics in a general scheme
that can be applied to any continuous phase transition. One of the
most natural such schemes is to have the order parameter feed back on
the tuning parameter
\cite{FraysseSornetteSornette1993,DickmanETAL2000}. Indeed, such
coupling can force an approach to the critical point as the linear
system size $L$ diverges~\cite{PruessnerPeters2006}, and in sandpile
models a narrative of such a coupling seems natural: Driving the
system increases the tuning parameter until the critical point,
$\zeta_c$, is reached and activity ensues. Activity then leads to
diffusion-like motion of particles through the system and to
dissipation at the boundaries, that is, a reduction of the tuning
parameter to below its critical value.

This general scheme can be summarized by the equation of motion
\cite{DickmanVespignaniZapperi1998}
\begin{equation}
\partial_t \zeta = h - \epsilon \rho_a(\zeta,t;L),
\elabel{1}
\end{equation}
where $h$ is a driving rate and $\epsilon$ represents a
coarse-grained, or bulk, dissipation. The activity, $\rho_a$, depends
on time $t$, and is treated as a noise term. The average order
parameter $\ave{\rho_a}=\frac{h}{\epsilon}$ is readily obtained from
the stationary state of this equation.  As long as $\ave{\rho_a}$
vanishes in the thermodynamic limit, $L\to \infty$, criticality can be
reached.  However, the description of SOC as a result of a linear
coupling between order and tuning parameter does not constrain the
dynamics sufficiently to explain the identity of scaling behaviors. In
this scheme (unlike in sandpile models), na\"ive finite-size scaling
analyses of thermodynamic observables do not show the universal
scaling exponents. By ``na\"ive'' we mean straight-forward numerical
measurements of observables like $\ave{\rho_a}(L)$, as performed in
SOC sandpiles, without taking into account (or knowledge of) the
scaling of the driving and dissipation, see below.  To na\"ively
observe universal finite-size scaling without having to account for
the details of the tuning dynamics it would be necessary that
$\ave{\rho_a(L)}\propto L^{-\beta/\nu}$, where $\beta$ and $\nu$ are
the exponents of the underlying phase transition. This can only be
true if the system-size dependence of $h$ and $\epsilon$ (introduced
by the facts that the density of dissipating boundary sites vanishes
as $L^{-1}$ and avalanche durations increase with $L$ such that
$\lim_{L \to \infty} h(L) =0$) is such that $h(L)/\epsilon(L)\propto
L^{-\beta/\nu}$.  In the coarse-grained description of coupled order
and tuning parameters there is no reason why this last condition
should be fulfilled.  Assuming power laws $h(L)=h_0 L^{-\kappa}$ and
$\epsilon(L)=\epsilon_0 L^{-\omega}$, we obtain scaling but not with
the universal exponents (assuming any functional form other than power
laws produces no scaling at all). There is an undetermined parameter,
$\kappa-\omega$, which must be chosen by design to force the universal
finite-size scaling exponents of the underlying phase transition upon
the self-tuning system.  Paradoxically, the scheme produces a system
whose tuning parameter does assume its critical value in the
thermodynamic limit, yet the exponents derived from na\"ive
finite-size scaling analyses are non-universal. This problem was
described in detail in ref.~\cite{PruessnerPeters2006}, and further
discussed in ref.~\cite{AlavaETAL2008,PruessnerPeters2008}.  Exponents
estimated at criticality, \eg from the spatial decay of correlation,
or from critical scaling need not be affected by an arbitrary choice
for $\kappa-\omega$.

In the present study we explore a scheme that achieves two goals: The
tuning parameter reaches the critical point in the thermodynamic
limit, and the exponents derived from na\"{i}ve finite-size scaling
analyses are identical to those of the underlying phase
transition. Our goal is to understand the interplay between
na\"{i}vely observed universality and the dynamics of self-tuning in
generic systems.  Here we choose a near-equilibrium system, but the
findings may also be informative for far-from-equilibrium systems,
like sandpile models.

The problem with coupling the tuning parameter to the order parameter
can be understood as follows. The system adjusts its tuning parameter
according to \eref{1} to achieve, in the thermodynamic limit, the
order parameter zero. For finite systems, however, a non-zero value is
targeted, which is arbitrary to the extent that $h(L)/\epsilon(L)$, or
equivalently $\kappa-\omega$, is arbitrary. The finite value is
necessary because at the level of description of \eref{1} fluctuations
in $\rho_a(t;L)$ ensure that $\ave{\rho_a}(L)$ is always finite, even
for $\zeta=0$, for finite systems. Targeting zero (setting
$h/\epsilon=0$) would drive the system to zero tuning parameter and
make SOC impossible. It may be argued that this constitutes a problem
in the level of description and that in sandpile models the value
$\rho_a=0$ is reached in finite systems since the absorbing phase
(unlike the high-temperature phase in an Ising model) has no
fluctuations. Typically, however, the order parameter $\rho_a$, is
defined as the asymptotic (long time average) value of the density of
active sites, conditioned on the existence of active sites, see \eg
~\cite{DickmanETAL2001}. It is therefore always non-zero. In other
words a ``complete descripition'' would require a re-definition of
$\rho_a$ and a revision of the entire formalism developed so far. Our
approach is independent of the presence of fluctuations in the phase
of vanishing order parameter.

Tuning the order parameter to zero as $L$ diverges leaves an undesired
arbitrariness in the intermediate values. Instead, we use the most
prominent signals of criticality, \ie the critical singularities (\eg
susceptibility, heat capacity, correlation length). The advantage of
coupling the tuning parameter to such a singularity is that
na\"ive finite-size scaling is universal, as numerical
evidence suggests to be the case in SOC. This will be shown below.

For example we could couple the temperature $T$ of a magnetic
system to the susceptibility, $\chi(T;L)$ which diverges uniquely at
the critical point, $\chi(T;L\to \infty) \propto \abs{T \to
T_c}^{-\gamma}$. This could be described by an equation of motion for
the temperature
\begin{equation}
\partial_t T=k \partial_T \chi(T,t;L),
\elabel{3}
\end{equation}
where $k$ is a constant that is related to the relaxation time and
needs to vanish in the thermodynamic limit to prevent
destabilization. Under sufficiently slow dynamics, the stationary
temperature of \eref{3} will be that of maximum susceptibility, where
$\partial_T \chi(T;L)=0$. As is well known, the position in $T$ of the
peak of the susceptibility, $T_{\chi_{\text{max}}}(L)$, approaches the
critical point as $\abs{T_{\chi_{\text{max}}}(L)-T_c} \propto
L^{-1/\nu}$. A derivation of this result from scaling arguments based
on the renormalization group can be found in ref.~\cite{Cardy1996},
p.~72. The prefactor determining the shift in
$T_{\chi_{\text{max}}}(L)$ for finite $L$ depends on the boundary
conditions of the system, but for our argument only the scaling
behavior is important. If the system is well characterized by the
average temperature it assumes under these dynamics, $\ave{T}$, it
will be equivalent to a system approaching $T_c$ with the exponent
needed to na\"ively observe the universal scaling behaviour, \ie that
of a system fixed at temperature $T=T_c$, while $L$ diverges. In Monte
Carlo studies of systems with unknown critical tuning parameter
values, \eg ref.~\cite{HsiaoMonceauPerreau2000}, similar arguments
ensure that scaling exponents can be derived numerically.

With $T(L)-T_c \propto L^{-1/\nu}$, we find $\ave{|m|}(T(L))\propto
(T(L)-T_c)^{\beta} \propto L^{-\beta/\nu}$, which is the universal
scaling behavior. The same argument holds for all other thermodynamic
observables, and equivalent results are obtained if we couple to any
other quantity that diverges uniquely at $T_c$. Thus, by coupling the
tuning parameter to the susceptibility the na\"ive finite-size scaling
becomes fully universal.

The above ideas are implemented by modifying a 2-$d$ Ising model and
allowing it to adjust its dimensionless coupling constant,
$K=J/(k_BT)$, where $J$ is a ferromagnetic Ising coupling, and $k_B$
is Boltzmann's constant. Physically, this corresponds to adjusting the
temperature. We want to design a system that finds the maximum of the
susceptibility with a precision that only depends on one scale, which
may be linked to the system size. Instead of measuring the first
derivative $\partial_{T} \chi$ required in \eref{3} we choose the more
robust method of bracketing to approach the maximum, see \eg
ref.\cite{PressETAL2002}, Chapter 9. To this end, three systems of
equal size and shape but different initial temperatures $K_1<K_2<K_3$
are run simultaneously for $t_{\text{max}}$ Monte Carlo steps. The
first half of this time is used for equilibration; in the second half
the first and second moments of the magnetization are recorded, and
from them the susceptibilities $\chi(K_i)$ are calculated.

Ordering the temperatures by the corresponding susceptibilities, there
are now $3!=6$ possible scenarios. These can be grouped into two
cases:

If the ordering of temperatures indicates that the maximum has not
been bracketed, \ie $K_{\chi_\text{max}}>K_3$ or
$K_{\chi_\text{max}}<K_1$, then the search range is widened by
reassigning the temperature furthest away from
$T_{\chi_\text{max}}$. For example, if $K_{\chi_\text{max}}>K_3$, then
$K_1$ is shifted to $K'_1=K_3 \sqrt{K_2 K_3}/K_1$.  Other choices are
possible, \eg $(K_3)^2/K_1$; the only requirements are that the
reassignment do not introduce any special scales (as would be the case
\eg for the choice $K_3+\Delta K$ with a fixed $\Delta K$) and that it
widen the search range. The choice made here is convenient for its
numerical stability.

If the maximum appears to be bracketed, \ie
$K_1<K_{\chi_{\text{max}}}<K_3$, the search range is narrowed by
halving the distance in $\log$-space between the middle-temperature
and the temperature furthest away from it. For example, if $K_3$ is to
be reassigned, it is shifted to $K'_3=\sqrt{K_2 K_3}$

Iterating this method the system (consisting of three Ising models)
will shrink its search range until the accuracy with which $\chi$ is
estimated during $t_{\text{max}}$ Monte Carlo steps forces it to widen
the search range. Due to fluctuations and the finiteness of
$t_{\text{max}}$, the measured $\chi(K_i)$ remain stochastic
variables, and the three temperatures will fluctuate around the true
$T_{\chi_\text{max}}$ for any given system size as exemplified in
\fref{1}.
\begin{figure}
\includegraphics[width=0.50\textwidth]{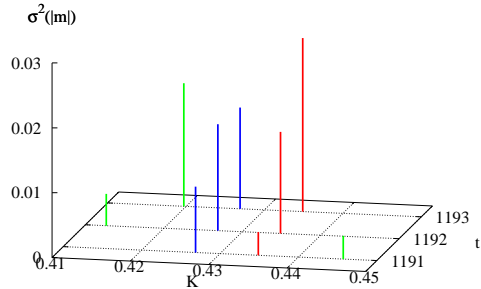}
\caption{Self-organizing couplings in a system of size $L=64$.  
Each time step consists of an equilibration of 388 scans over
the lattice and another 388 scans during which the susceptibility is
determined. At $t=1191$, the maximum is not bracketed and the search
range is widened. In the next time step it is bracketed, and the range
shrinks. Initially three temperatures far from criticality
($K_c\approx 0.44068$) are chosen that bracket the maximum, $K_1=0.3,
K_2=0.5, K_3=0.6$. The systems converge near the temperature of
maximum susceptibility. Statistics are collected after 1000
equilibration time steps.
}  \flabel{1}
\end{figure}
Importantly, however, the range of these fluctuations can be made
arbitrarily small by increasing $t_{\text{max}}$, see \fref{2}. As $L$
increases, the most likely and average temperatures approach $T_c$,
while the distributions 
become narrower.
\begin{figure}
\includegraphics[width=0.38\textwidth,angle=-90]{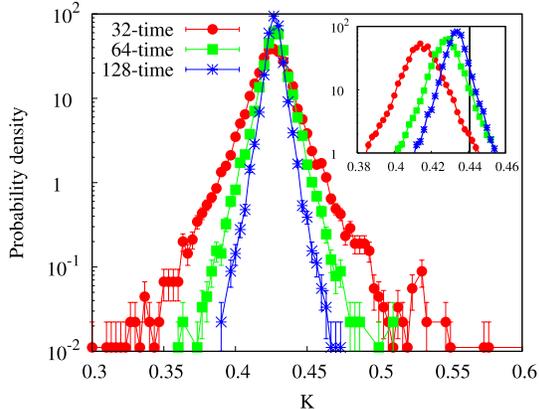}
\caption{The distribution of self-organized temperatures, can be made
arbitrarily narrow by increasing $t_{\text{max}}$, see main graph,
where $t_{\text{max}}=0.4 \times 32^{2.15}, 0.4 \times 64^{2.15},
0.4\times 128^{2.15}, $ for fixed $L=32$. Inset: Temperature
distributions for $L=16, 32, 64$, where $K$ is adjusted after a fixed
$t_{\text{max}}=0.1 \times 128^{2.15}$.}  \flabel{2}
\end{figure}
Holding $t_{\text{max}}$ fixed, one observes the expected finite-size
scaling of the temperature, \ie $|\ave{T}-T_c| \propto L^{-1/\nu}$ and
all other thermodynamic observables, \eg $\ave{\abs{m}} \propto
L^{-\beta/\nu}$, up to a certain system size where the accuracy with
which $T_{\chi_\text{max}}$ is estimated becomes insufficient.
Eventually the system becomes unstable as the assumption
$t_{\text{max}} \gg $ equilibration time becomes invalid.  In order
for na\"ive finite size scaling to be universal without a bound on
$L$, we require that $t_{\text{max}}$ increase sufficiently fast.  The
exact minimum speed at which $t_{\text{max}}$ must diverge with $L$
depends on the chosen dynamics and is given by the dynamical exponent
$z$. We require
\begin{equation}
t_{\text{max}} \propto L^x,
\elabel{t_max}
\end{equation}
where $x \geq z$. This was only superficially confirmed by estimating
the maximum equilibration times that destabilized systems of different
sizes. In \fref{3} numerical results of the most prominent observables
are presented for system sizes up to $L=512$, where $x=2.15$ was
used. For the single-spin flip dynamics used here, $z=2.13(3)$
\cite{Williams1985}.  Stricly speaking, the macroscopic time referred
to in \fref{1} to estimate the position of the peak of the temperature
distribution also has to increase with $L$. In the present study we
are always far from the limit where this issue becomes important.  Our
situation is fundamentally different from that of the order parameter
coupled to the tuning parameter, where there was only one special
choice for the exponent $\kappa-\omega$ that reproduced the known
finite-size scaling exponents. The new scheme is more robust since the
exponent $x \geq z$ is only restricted to a semi-infinite range.
\begin{figure}
\vspace{-2cm}
\scalebox{1.4}{\noindent\includegraphics[width=12pc,angle=0]{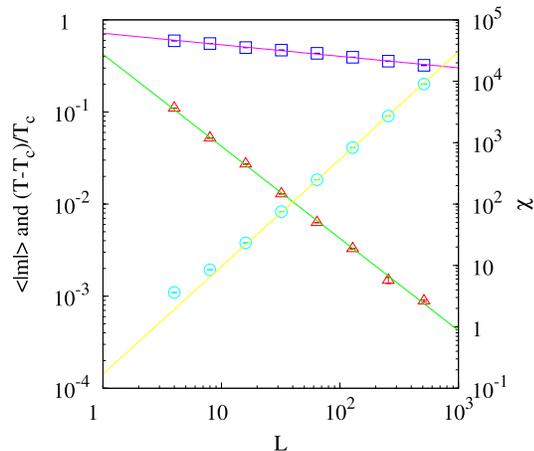}}
\vspace{-2cm}
\caption{The average self-organized temperature (triangles),
  magnetization (squares), and susceptibility (circles) in systems
  ranging from $L=4, 8, 16... 512$. Straight lines represent power
  laws with the known exponents of the 2-d Ising universality class,
  symbols are measurements.}  \flabel{3}
\end{figure}
This situation is very similar to sandpiles in the following sense:
For a sandpile to display proper scaling for all system sizes, the
intervals (measured in a microscopic time scale of individual
topplings) between additions of grains must diverge with system size
fast enough. However, it is impossible to drive a sandpile ``too
slowly''. If the drive is slower than necessary, then the sandpile
will be inactive for a while between avalanches, but the avalanches
and associated order-parameter properties will still obey the
universal scaling laws.

The idea for the present study emerged from a discussion of biological
evolutionary systems that are believed to maximize a form of
susceptibility, as they balance the need for rigidity to store
information (for example in the form of DNA) against the need for
flexibility to respond to new situations. While the use of multiple
copies of a system may seem curious in the context of sandpiles, it
is unavoidable in the context of evolution.

The correlation length, $\xi$, which also diverges at criticality,
seems the most natural means to discuss the relevance of the proposed
mechanism to sandpile models. It has been speculated that the
correlation length plays an important role in the feedback between
order- and tuning parameter \cite{FraysseSornetteSornette1993}.  The
correlation length measures the spatial distance over which
perturbations to a system can be communicated. In thermal systems
$\xi$ needs to be inferred from the spatial decay of correlation
functions. In sandpiles, on the other hand, the length over which
perturbations are communicated is dictated directly by the
dynamics. Any particle added to a sandpile must be transported to the
boundaries to be dissipated. The assumption of stationarity thus
implies that perturbations can be felt across the entire system, \ie
$\xi \propto L$. Clearly, this corresponds to the maximization of
$\xi$ for any $L$. Imposing this relation, just like maximizing the
susceptibility, leads to the preservation of na\"ively observed
universality, see also \cite{PruessnerPeters2006}. Therefore, thinking
of sandpile models as correlation-maximizers rather than
activity-minimizers,
observations of the universal scaling exponents 
become the expectation rather than the surprise.

In conclusion, a mechanism has been investigated which does more than
bring a system to the critical point. It reproduces the scaling
exponents observed in finite-size scaling studies where a
corresponding model is fixed at criticality. Thus the mechanism
preserves universality in na\"{i}vely performed finite-size scaling
analyses under conditions of self-tuning. This was achieved by letting
the tuning parameter maximize a quantity that diverges uniquely at
criticality. In contrast to a coupling between order and tuning
parameter, this allows us to use the well-defined maximum rather than
an arbitrary small parameter for orientation. The Ising model was used
to show that the process is indeed capable of recovering both the
well-known scaling exponents and the critical temperature.

\bibliographystyle{plain}

\end{sloppypar}
\enddocument